# Refining Coarse-Grained Molecular Topologies: A Bayesian Optimization Approach

*Pranoy Ray[1,2,3], Adam P. Generale[1,3], Nikhith Vankireddy[3,4], Yuichiro Asoma[5], Masataka Nakauchi[5], Haein Lee[5], Katsuhisa Yoshida[5], Yoshishige Okuno[5], Surya R. Kalidindi[1,2,3,*]*

[1] George W. Woodruff School of Mechanical Engineering, Georgia Institute of Technology, Atlanta, USA
[2] School of Computational Science and Engineering, Georgia Institute of Technology, Atlanta, USA
[3] Multiscale Technologies Inc., Seattle, USA
[4] Purdue University, West Lafayette, USA
[5] Resonac Corporation, Tokyo, Japan

*Corresponding author email: surya.kalidindi@me.gatech.edu

## ABSTRACT

Molecular Dynamics (MD) simulations are vital for predicting the physical and chemical properties of molecular systems across various ensembles. While All-Atom (AA) MD provides high accuracy, its computational cost has spurred the development of Coarse-Grained MD (CGMD), which simplifies molecular structures into representative beads to reduce expense but sacrifice precision. CGMD methods like Martini3, calibrated against experimental data, generalize well across molecular classes but often fail to meet the accuracy demands of domain-specific applications. This work introduces a Bayesian Optimization-based approach to refine Martini3 topologies - specifically the bonded interaction parameters within a given coarse-grained mapping - for specialized applications, ensuring accuracy and efficiency. The resulting optimized CG potential accommodates any degree of polymerization, offering accuracy comparable to AA simulations while retaining the computational speed of CGMD. By bridging the gap between efficiency and accuracy, this method advances multiscale molecular simulations, enabling cost-effective molecular discovery for diverse scientific and technological fields.

## INTRODUCTION

Coarse-grained molecular dynamics (CGMD)[1,2] has emerged as a vital tool for material development, offering crucial insights into complex molecular systems including polymers[3], proteins[4], and membranes[5]. The primary advantage of CGMD is its ability to explore molecular phenomena over larger length scales and longer time frames, surpassing the capabilities of traditional all-atom molecular dynamics (AAMD)[6–11] simulations, which typically offer higher resolution and, hence, are particularly adept at capturing detailed interfacial interactions[12]. In detail, CGMD achieves this speedup by effectively representing groups of atoms as beads[13–18], thus extending the simulation capabilities from picoseconds to microseconds temporally and from nanometers to micrometers spatially. Consequently, coarse-graining provides unprecedented insights into complex molecular phenomena that remain inaccessible to conventional AAMD, thus enabling the study of complicated phenomena such as the self-assembly behaviors of polymers[19].

Emergent CGMD modeling toolsets rely on two key components to learn the underlying inter-molecular relationships: bead-mapping schemes and the parametrization of bead-bead interactions.

In this work, 'molecular topology' specifically refers to the set of bonded parameters (bond lengths, angles, and their associated force constants which elucidate a molecule's topology) defined within a given coarse-grained mapping, rather than to the optimization of the bead-mapping scheme itself. These components are developed using two primary approaches: top-down[13–15] and bottom-up[16–18]. Top-down approaches simplify systems to reproduce macroscopic properties with frameworks like CG-Martini[13,20–22], where up to four heavy atoms are mapped onto one bead, and, the inter-bead interactions are parametrized using experimentally obtained thermodynamic data. In particular, Martini 3[21–23], the most recent version of the CG-Martini force field, typically offers reasonable coarse-approximated accuracy when widely applied across biological and material systems[22,24] but struggles with materials exhibiting varying degrees of polymerization. Conversely, the bottom-up approach derives parameters directly from all-atom molecular dynamics (AAMD), ensuring microscopic accuracy but often requiring computationally expensive iterative refinement to match target observables. While top-down strategies aim to reproduce macroscopic properties and offer broader applicability, bottom-up methods emphasize fidelity to atomistic interactions. Recent advances in machine learning increasingly automate parameterization—particularly in polymer systems—making the choice between these approaches contingent on system complexity, desired accuracy, and modeling objectives.

Over the past decade, machine learning (ML)[25,26] has transformed coarse-grained (CG) mapping and parameterization processes, markedly improving the accuracy and efficiency of CGMD simulations. In particular, ML-driven CGMD approaches leverage advanced algorithms to extract or optimize target parameters from large datasets, while also enabling active learning workflows that iteratively refine models. These methods are especially well-suited for bottom-up methodologies reliant on AAMD data. The relative computational affordability and accessibility of AAMD simulations, compared to experimental measurements, facilitate not only the generation of high-quality training datasets but also the on-demand data acquisition required for active learning, ensuring models remain adaptive and robust. Notable advancements in this pursuit include the Versatile Object-oriented Toolkit for Coarse-graining Applications (VOTCA)[27], which integrates techniques such as Iterative Boltzmann Inversion, force matching, and Inverse Monte Carlo. Similarly, the software MagiC[28] implements a Metropolis Monte Carlo method, providing enhanced robustness against singular parameter values during optimization. Emerging ML-driven frameworks further generalize these capabilities: chemtrain[29] enables learning deep potential models via automatic differentiation and statistical physics, while DMFF[30] provides an open-source platform for differentiable force field development with support for both top-down and bottom-up approaches. Tools like TorchMD-Net[31] and DeepMD[32] enable end-to-end differentiable force field training with built-in uncertainty quantification, extending these principles to coarse-grained systems. Beyond these, other methods[33] employing Relative Entropy Minimization have been integrated as well with popular simulation engines like GROMACS[34] and LAMMPS[35]. For small molecules, approaches based on partition functions[36] and parameter tuning via quantum chemical calculations[37] instead of directly running AAMD have also shown sufficient promise. In this regard, optimization algorithms are central to resolving complex challenges in CG force field development because they systematically explore high-dimensional parameter spaces to minimize discrepancies between coarse-grained and reference data, ensuring accurate and transferable models while addressing the inherent nonlinearity and complexity of molecular interactions. Amongst the optimization methods, gradient-based techniques and Evolutionary Algorithms (EAs)—notably Genetic Algorithm (GA) and Particle Swarm Optimization (PSO)—have gained prominence. GA has been applied to optimize parameters in ReaxFF reactive force fields[38] and coarse-grained water models[39], while PSO has been used in tools like Swarm-CG[40] and CGCompiler[41] for CG model parameterization. However, while EAs are effective in exploring vast parameter spaces, they can be computationally expensive and often require numerous evaluations of the objective function. While gradient-based methods[42,43] excel in problems with smooth, differentiable objective functions, they face limitations in CG force field parameterization due to (i) the non-differentiable nature of MD simulation outputs and (ii) their propensity to converge to local minima in complex energy landscapes. In this regard, Bayesian Optimization (BO)[44–47] offers a powerful approach for problems where objective function evaluations are expensive and data acquisition is costly. By balancing exploration and exploitation through a probabilistic model, BO efficiently converges to optimal solutions with fewer evaluations, making it well-suited for optimizing CG force field parameters where computational cost is critical. Its ability to incorporate prior knowledge and handle noisy or sparse data further enhances its applicability to force field optimization tasks. Recent studies highlight BO's advantages in diverse contexts. For example, BO optimized chlorine dosing schedules for water distribution systems[48] with significantly fewer evaluations than traditional EAs. In materials design, BO identified superior solutions with smaller sample sizes and fewer iterations compared to GA and PSO[49]. Additionally, BO-based methods[50] have demonstrated strong performance in problems with small dimensions and limited evaluation budgets. These findings underscore BO's suitability for CGMD parameter optimization in polymer systems, particularly for computationally expensive simulations involving higher degrees of polymerization[51,52]. Furthermore, previous studies[53,54] have highlighted the need to re-parametrize Martini for specific systems, including MOFs, proteins, and polymers, to address its limitations in accuracy for certain applications. In summary, while the conventional approach involves optimizing parameters at lower degrees of polymerization and validating at higher degrees, effectively addressing mesoscale phenomena necessitates models that are capable of systematically accounting for and adapting to variations in the degree of polymerization. BO's advantages over gradient-based methods are particularly pronounced in high-dimensional CG parameter spaces. Although gradient-based optimization scales formally to higher dimensions, it requires precise derivative calculations—a significant challenge when objectives involve computationally expensive MD simulations with inherent noise. BO circumvents this by treating the optimization as a black-box problem, strategically balancing exploration and exploitation through probabilistic surrogate models. This enables global optimization without requiring gradient information, making it robust to noisy evaluations and better suited for identifying transferable parameter sets across polymerization degrees.

Martini3 has excelled as a general-purpose tool for the baseline coarse-graining of wide-ranging molecules. Detrimentally, its generality also precludes its ability to provide accurate property predictions for any particular sub-classes of molecules. The approach presented in this work aims to directly remedy this limitation by a low-cost protocol based on active learning for the efficient low-dimensional parametrization of the bonded parameters of a CG molecular structure. Particularly, we use BO to enhance the accuracy of the Martini3 force fields for three common polymers across varying degrees of polymerization. The corresponding AAMD calculation results are defined to be the ground truth in our iterative refinement scheme. We calibrate the BO model on density ($\rho$) and radius of gyration ($R_g$) and demonstrate a unique generalizable parametrization scheme for CG force field optimization, independent of the degree of polymerization. Furthermore, because this BO framework optimizes against abstract target properties, it is inherently flexible and can be readily adapted to calibrate CG models against experimental macroscopic data, in addition to the AAMD-based refinement demonstrated here.

**RESULTS AND DISCUSSION**

**Low-dimensional Parametrization of the Coarse-grained Molecular Structure**

Molecular Dynamics (MD) models define the geometry of molecular topology through the bonded parameters of the force field such as bond lengths ($b_0$), bond constants ($k_b$), angle magnitudes ($\Phi$), and angle constants ($k_\Phi$). These topological parameters are intrinsically linked to macroscopic properties of molecules, including density ($\rho$) and radius of gyration ($R_g$). Variations in these topological parameters directly influence molecular geometry, which in turn alters packing efficiency, spatial distribution of atoms/beads, and overall molecular compactness. For instance, increasing bond lengths ($b_0$) or widening bond angles ($\Phi$) typically leads to a larger molecular volume, which impacts bulk properties such as $\rho$, while bond constants ($k_b$) and angle constants ($k_\Phi$) reflect the stiffness of the polymer backbone, which is crucial for determining conformational properties like $R_g$. Additionally, polymer chains containing aromatic rings introduce an additional bond length parameter ($c$) to account for the constant aromatic bonds needed to preserve the ring's topology. Therefore, this set of bonded parameters ($\theta$) can be defined as:

$$\boldsymbol{\theta} = \begin{cases} [b_0, k_b, \Phi, k_\Phi] & \text{for non-aromatic molecules} \\ [b_0, k_b, \Phi, k_\Phi, c] & \text{for aromatic molecules} \end{cases} \quad (1)$$

Notably, this study excludes dihedral angles from the parameter set due to the complexity of their conformational space and the non-trivial relationship with the aforementioned macroscopic properties. However, the number of topological parameters ($\theta$) scale linearly with the degree of polymerization ($n$), hence attempting to optimize every parameter within the molecular topology space would be highly computationally inefficient, even in the case of CG representations. Consequently, reducing the dimensionality of the parameter space becomes essential. This reduction in the number of parameters to be optimized reduces the number of design variables, thus enhancing the efficiency and tractability of Bayesian optimization (BO), which is typically less effective in higher-dimensional spaces. To this end, we propose an effective low-dimensional parametrization of a CG molecule's topology, which focuses on capturing the critical degrees of freedom that influence vital macroscopic properties like $\rho$ and $R_g$. In particular, we consider the bonded parameters ($\theta$) of the first, middle, and end bonds of a polymer chain to sufficiently describe the CG molecular topology. This consideration enables an efficient CG topology representation because the middle bonds capture the essence of the uniform internal structure, while the first and last bonds capture the deviations due to functional groups or termination effects. This selection is deemed adequate, given the regularity of repeating units in a polymer chain, and unique boundary effects at chain ends. The rationale behind this low-dimensional approach, distinguishing parameters for start, middle, and end segments, is to capture transferable features. The parameters optimized for these regions are intended to be applicable when constructing models for other degrees of polymerization of the same polymer, thus promoting transferability. A direct comparison to a model using a single, uniform parameter set for all segments was not performed but would be valuable future work to precisely quantify the impact of modeling these terminal boundary effects. Furthermore, this low-dimensional parametrization approach balances computational efficiency with the need to capture key topological features of the CG polymer.

Furthermore, this low-dimensional parametrization approach balances computational efficiency with the need to capture key topological features of the CG polymer. For instance, a coarse-grained styrene monomer contains 5 bonded parameters (see Supplementary Figure 1) while a 20-polystyrene polymer chain (20-PS) consists of 139 bonded parameters (see Figure 1). However, if we consider the start, middle, and end bonds of 20-PS (which contains aromatic rings in every monomer), we arrive at 3 sets of $\theta$, which gives us 15 dimensions to optimize (see Figure 1).

**CG Topology Optimization Framework**

Leveraging the strengths of CGMD simulations and Bayesian optimization (BO) in an integrated manner, addresses a critical challenge: improving the fidelity of Martini3 forcefields while maintaining computational efficiency. To this end, we create a synergistic workflow (see Figure 2) that optimizes the pre-selected topological parameters ($\theta$) for a particular polymer chain, by calibrating on the CGMD-derived macroscopic properties of the bulk polymer. We initiated the workflow with 20 CGMD simulation runs, based on topological parameters ($\theta$) selected with a space-filling Latin Hypercube experimental design[55] with maximum projection. This initial dataset was used to train the Gaussian Process (GP) surrogate model, ensuring adequate coverage of the parameter space. Following this, we integrated the Martini3 simulation method with Bayesian Optimization (BO), leveraging the Expected Hypervolume Improvement (EHVI) acquisition function[56]. The optimization proceeded iteratively over ~50 iterations, with two CGMD simulations conducted per iteration. Model convergence was defined as a plateau in loss reduction (i.e., objective function), indicating that the optimization had identified an optimal set of parameters ($\theta_{optimal}$). This dual-run strategy balances computational efficiency with the need for sufficient data to refine the GP model, thus enabling effective exploration with the exploitation of the parameter space. Resultantly, the predicted target estimates are achieved by minimizing the objective function's loss, which can be defined as:

$$\widehat{\boldsymbol{\theta}} = \operatorname*{argmin}_{\theta} \|k^{CG}(\boldsymbol{\theta}) - k^{AA}\|_2 \qquad (2)$$

where the topological parameters ($\theta$) serve as input variables, and the properties derived from CGMD runs ($k_{CG}$) are compared against reference high-fidelity data (AAMD-derived property estimates - $k_{AA}$). By directly targeting macroscopic properties such as density and radius of gyration, this approach is designed to ensure fidelity in key emergent behaviors and avoid potential divergences in these properties that can occur in bottom-up methods focused primarily on matching local structural distributions (e.g., radial distribution functions). In summary, by combining this approach for iterative refinement, we facilitate multi-objective Bayesian optimization (MOBO), by minimizing the objective function while maintaining computational feasibility. Table 1 defines the optimization space (constraints) for these $\theta$. This parametrization can be extended to complex polymer systems by introducing additional parameters while consistently using an adequately low number of topological parameters.

**Pareto Optimal Property Discrepancy Frontier**

The polymer systems trained in this work exhibit significant structural and chemical diversity across four degrees of polymerization, which necessitates varying training times even when trained on the same computational resource. To this end, we train the integrated CGMD-MOBO models individually on a single NVIDIA A100 GPU with 32 GB of memory, and the training times average between 3 hours to 11 hours (depending on the CGMD simulation times for a bulk polymer chain system). The models are trained until the objective function plateaus (see Figure 3), indicating that the total loss has been minimized within the optimization search space (as shown in Table 1). Specifically, every polymer's Martini3 topology is optimized until the absolute percentage errors converge to less than ~10%. Through Figure 3, we also observe that the radius of gyration ($R_g$) as a property facilitates quicker convergence, while the interdependencies affecting the density ($\rho$) demand additional iterations to balance competing influences. These interdependencies are attributed to the complexity of their optimization landscapes. Specifically, since $R_g$ is a molecular-scale property, it is directly influenced by localized geometry-based structural parameters such as bond lengths and angles, which are simpler to optimize and exhibit a smoother landscape. In contrast, $\rho$, a bulk property, is affected by both bonded and non-bonded interactions, which requires adjustments to long-range interactions (non-bonded parameters) and packing efficiency (bonded parameters). This leads to a more complex and slower-converging optimization process. Conversely, it can be claimed that the initial parameter set may have been closer to the optimal region $R_g$, further accelerating its convergence, hence a bias. However, to counter that thought, the model for every polymer chain was re-initialized and re-trained over 5 seeds, owing to the stochastic selection process of BO. For every seed, it was observed that $R_g$ yielded quicker convergence, hence supporting our prior notion that $\rho$ depends on non-bonded interactions as well.

Convergence to Pareto optimal values typically occurred within ~50 iterations (100 CGMD simulation runs), as shown in Figure 3. The evolution of the convex Pareto front (see Figure 4) over these 50 iterations represents the set of non-dominated solutions, where improvements in one objective (density or radius of gyration) cannot be made without degrading the other. In particular, the front effectively captures the trade-offs between these competing objectives, allowing for informed decision-making in selecting optimal CG parameters. For instance, Figure 4 shows the convex Pareto fronts we achieve with our workflow for 10-PE and 50-PE. The progression of the color bar denotes the evolution of the front cover 60 iterations, including 10 initial iterations (with 20 CGMD runs). This demonstrates that the integration of CGMD with MOBO effectively navigates the parameter space to balance fidelity and computational cost, by strategically exploiting known optimal regions while systematically exploring uncertain areas.

The improvements in the relative efficiency of our integrated framework are further quantified by conducting a standard deviation analysis to compare our optimized CG topologies to the raw Martini3 topologies, over 5 seeds per polymer. Figure 5 shows the percentage error metrics for density ($\rho$) and radius of gyration ($R_g$) across the polymers of interest. This figure clearly illustrates the superior performance of our optimized topologies (green line) as compared to the raw Martini3 topology (red line). This MOBO solution also demonstrates the generalizability and superiority of our proposed framework, by efficiently reducing discrepancies between CGMD and AAMD from a bottom-up perspective.

For PE and PS, the absolute percentage error in ρ decreases by ~30% as n increases from 3 to 50 (Figure 5, top). However, PMMA exhibits increased errors at higher degrees of polymerization due to Martini3's limited representation of polar-nonpolar bead interactions in elongated chains.[23] This aligns with prior studies[57] showing that force field accuracy for polar polymers degrades with chain length when non-bonded parameterizations are insufficient. Moreover, the optimized force field consistently reduces errors for ρ and $R_g$ across all polymer families. For instance, in 3-PE, coarse-graining three atomic groups into a single bead in the Martini3 model leads to deviations in molecular behavior and conformational flexibility. The proposed framework adjusts interaction parameters, improving the force field's fidelity. On the other hand, in PMMA, the Martini3 model demonstrates higher accuracy at low polymerization degrees due to the effective separation of polar and non-polar blocks within beads. This nature of coarse-graining limits flexibility at higher molecular weights, increasing error. However, in PS, significant errors in density are linked to benzene ring stacking, which is inadequately represented by Martini3. Interestingly, the original Martini3 model aligns better with experimental densities than the optimized version for 50-PS, highlighting the need to balance future optimization efforts towards both experimental results (high-fidelity) and AAMD data (low-fidelity).

This paper introduces a MOBO (Multi-Objective Bayesian Optimization) framework that significantly enhances CG-Martini3 topologies for common polymers like PE, PMMA and PS. Our findings challenge conventional raw Martini3 topologies by consistently yielding CG topologies that accurately reproduce AA-calculated macroscopic properties (density and radius of gyration), with improvements observed across all studied materials and degrees of polymerization. Furthermore, this work presents a framework for the low dimensional parametrization of CG molecular topologies, which increases its generalizability over unknown complex polymer systems. A key strength of the proposed low-dimensional parametrization is its design for transferability across varying degrees of polymerization, moving towards a unified model for a given polymer type. The framework reports exceptionally low errors under ~10% for both density and radius of gyration for all the 12 polymers studied and introduces itself as a new benchmark for future model-building efforts. This unprecedented improvement helps bridge the gap between low- and high-fidelity MD models, enabling accurate predictions with CGMD at a fraction of the corresponding AAMD's computational expense.

## METHODS

### Molecular Dynamics Simulations

Molecular dynamics (MD)[3,58,59] simulations scale in computational time and complexity with the increase in the number of particles simulated. In general, AAMD simulations are expensive and time-consuming when we tend to simulate thousands of atoms. Because of the larger number of atoms simulated, the greater number of physical interactions between them need to be captured to provide a realistic representation of the changes in the system under specific temperature and pressure conditions. The fundamental challenge in running AAMD simulations is the high computational budget necessary to simulate the system, which is often infeasible and, hence, a bottleneck. However, with the advent of CGMD as a methodology, MD simulations have grown to be more scalable for molecular systems with a higher number of atoms. In fact, for the same molecule, AAMD simulations involve substantially higher computational costs as they simulate a higher number of particles (here, atoms - higher resolution), as compared to CGMD, which simulates a smaller number of particles (here, beads - lower resolution). Coarse graining (CG) is fundamentally aimed at simplifying complex systems, and in the context of all-atom structures, coarse-graining involves grouping multiple atoms into a single bead while retaining as much information as possible from the original structure and composition. There is always an information loss or approximation when one moves from an all-atom system towards a coarse-grained system. However, the goal is to accept the information loss to a reasonable extent in the pursuit of speeding up the simulation time by multiple orders of magnitude while also saving on computational cost.

Supplementary Figure 1 shows a schematic of the coarse-graining procedure for a poly-styrene monomer. Martini3[21–23] approximates aromatic rings (note the grey beads - three in number, which help to coarse-grain the aromatic ring) as well as molecular chains (the singular green bead, which helps with coarse-graining the carbon chain shown as $C_2H_4$).

**Polymers of Interest**

The polymers we focus on in this work include Polyethylene (PE), Polystyrene (PS), and Polymethyl Methacrylate (PMMA) across multiple degrees of polymerization ($n: n \in \{3,10,20,50\}$). The wide applicability of these polymers[60–62], combined with their reusability in building complex polymer systems, motivated our choice of these specific polymer systems. The all-atom (AA) molecular structure files (.gro/.pdb) for the aforementioned 12 polymer chains were prepared by J-OCTA software[63], with the bonded and non-bonded interactions being parametrized by the GAFF (General Amber Force Field)[64,65]. Necessary electrostatic potential charges for each atom were calculated using Gaussian16[66] revB.01 with RHF/6-31G(d) level of theory. Coarse-graining was performed on these all-atom molecular structures using Martini3[13,22] with the martinize2[67,68] and vermouth[67] python codebases. The Martini3 force field is generated in accordance with the pre-defined Martini Interaction Matrix[13], which contains four main types of interaction sites: Polar (P), nonpolar (N), apolar (C), and charged (Q). Within the main interaction sites, there are sub-levels, and the interactions between each sub-level across different interaction sites are captured in an interaction matrix[22] with LJ potential values assigned for each interaction. Despite the limited applicability of Martini on polymers, it serves as a great starting point for helping map AA to CG molecular structures and topologies. For instance, Table 2 shows an example per polymer of the AA to CG structure mapping for 20 degrees of polymerization.

**Molecular Dynamics simulation setup**

The AAMD calculations were conducted using the GROMACS[34] simulation package. The calculation of derived properties, such as the radius of gyration of polymer chains and the density, was also performed using GROMACS. We evaluate ensembles of 100 AA molecular structures for the 12 types of polymer chains. Three-dimensional periodic boundary conditions have been adopted for the simulation cell to place 100 polymer chains randomly. MD simulations were performed with a time step of 2 fs in the NPT ensemble, using the V-rescale thermostat (T = 300 K, P = 1 bar) and C-rescale barostat for 100 ns. The convergence of the radius of gyration of each polymer chain has been confirmed at ~10 ns; therefore, the rest of the run was used for sampling.

All CGMD calculations were performed with GROMACS with the Martini3 force field as defined earlier. We evaluate ensembles of 100 coarse-grained polymer chains of each polymer of interest, representing a baseline level of accuracy. These ensembles were subjected to energy minimization, followed by NVT equilibration and, finally, NPT equilibration for 10 ns, using the V-rescale thermostat (T = 300 K, P = 1 bar) and C-rescale barostat for 100 ns. The convergence of the radius of gyration of each polymer chain has been confirmed at ~8 ns; therefore, the rest of the run was used for sampling.

Supplementary Figure S2 evaluates the absolute error of raw Martini3 forcefield with respect to corresponding all-atom (AA) computations. We observe that the errors lie in the range of 20-30% for most cases, the maximum error being over 60%, which shows the inaccuracies posed by computing macroscopic properties from CGMD using the raw Martini3 force field.

**Bayesian Optimization (BO)**

Optimization of expensive underlying functions is a problem endemic to multiple scientific fields of study, including material informatics[69,70], bioinformatics[71], manufacturing[72], and economics[73]. Frequently, these costly predictive tools only enable the capability to query select points in the input space without any ability to differentiate with respect to the response - resulting in essentially "black-box" functions. Bayesian Optimization (BO) has emerged as an efficient method for optimizing such black-box functions $f$, formalized as

$$x^* = \underset{x \in \mathcal{X}}{argmax} f(x) \tag{3}$$

where $\mathcal{X}$ denotes the input space over which a solution is sought and $x^*$ the input maximizing $f$. The gain in efficiency towards addressing problems of this type is primarily due to the ability to exploit information theory[74] and Bayesian inference over the underlying function space, frequently through the creation of Gaussian process surrogates[75].

**Multi-Output Gaussian Process Regression**

Gaussian Processes ($\mathcal{GP}$s)[76] are widely used probabilistic surrogate models. In the context of Bayesian Optimization (BO), they are similarly leveraged towards approximating the true underlying objective function over which to optimize. ($\mathcal{GP}$s) can be viewed as probability distributions over function spaces, providing essential properties related to Bayesian analysis[77,78]. This relationship is denoted as a $\mathcal{GP}$, i.e., $f(\cdot) \sim \mathcal{GP}(v(\cdot), k(\cdot,\cdot'))$, which is uniquely determined through a mean function $v(\cdot)$ and a covariance function $k(\cdot,\cdot')$ parameterized by hyperparameters, $\boldsymbol{\theta}$. Often, the mean function is taken to be $v \equiv 0$ without loss of generality.

Given a training dataset $\{(\boldsymbol{x}_n, y_n)\}_{n=1}^{N}$ of $N$ corrupted observations with an assumed Gaussian noise $\xi_i \sim \mathcal{N}(0, \sigma_y^2)$, the collection of all training inputs can be denoted as $\boldsymbol{X} \in \mathcal{R}^{N \times M}$, the vector of all outputs as $\boldsymbol{y}$, and $\boldsymbol{f}$ the infinite-dimensional process latent function values. The particular covariance function used in this work is the automatic relevance determination squared exponential (ARD-SE)[76], defined as

$$k(\boldsymbol{x}, \boldsymbol{x}') = \sigma_f^2 \, exp\left(-\frac{1}{2} \sum_{m=1}^{M} \frac{(x-x\prime)^\top (x-x\prime)}{\lambda_m^2}\right) \tag{4}$$

where $\lambda_m$ is the lengthscale associated with input dimension $m$ of $M$, and $\sigma_f$ the amplitude. The resulting set of hyperparameters for this covariance function is then $\boldsymbol{\theta} = \{\boldsymbol{\lambda}, \sigma_f\}$. $K(\boldsymbol{X}, \boldsymbol{X}')$ represents the constructed covariance matrix using the covariance function established in Eq. (4). For legibility, this will be abbreviated as $\boldsymbol{K_{ff}}$ to denote the covariance matrix constructed with the available training dataset, defining the latent process. The hyperparameters of this covariance function are inferred through maximizing the log marginal likelihood.

$$logp(\boldsymbol{y}|\boldsymbol{X}, \boldsymbol{\theta}) = \frac{1}{2}\boldsymbol{y}^\top \boldsymbol{K}_y^{-1}\boldsymbol{y} - \frac{1}{2}\log|\boldsymbol{K}_y| - \frac{N}{2}\log 2\pi \quad (5)$$

where $K_y = K(\boldsymbol{X}, \boldsymbol{X}') + \sigma_y^2 \boldsymbol{I}$. Predictions of this base model can be expanded to handle multi-output functions in a similar manner to the scalar output case, through expansion of the covariance matrix to express correlations between related outputs[79]. Such Multioutput Gaussian processes (MOGP) learn a multioutput function $f(\boldsymbol{x}): \mathcal{X} \rightarrow \mathbb{R}^P$ with the input space $\mathcal{X} \in \mathbb{R}^D$. The p-th output of $f(\boldsymbol{x})$ is expressed as $f_p(\boldsymbol{x})$, with its complete representation given as $\hat{f} = \{f_p(\boldsymbol{x})\}_{i=1}^P$. MOGPs are similarly completely defined by their covariance function (assuming $v \equiv 0$), resulting in a covariance matrix $\boldsymbol{K} \in \mathbb{R}^{NP \times NP}$. In this work, the multi-output covariance matrix is constructed through the Linear Model of Coregionalization (LMC)[79,80]. This model represents a method of constructing the multi-output function from a linear transformation $W \in \mathbb{R}^{P \times L}$ of $L$ independent functions $g(\boldsymbol{x}) = \{g_l(\boldsymbol{x})\}_{l=1}^L$. Each function is constructed as an independent $\mathcal{GP}$, $g_l(\boldsymbol{x}) \sim \mathcal{GP}(0, k_l(\boldsymbol{x}, \boldsymbol{x}'))$, each with its own covariance function, resulting in the final expression $f(\boldsymbol{x}) = \boldsymbol{W}g(\boldsymbol{x})$. The multi-output covariance function described by this model is then expressed as:

$$k(\{\boldsymbol{x}, p\}, \{\boldsymbol{x}', p'\}) = \sum_{l=1}^L W_{pl}\, k_l(\boldsymbol{x}, \boldsymbol{x}')W_{p'l} \quad (6)$$

which can be seen to encode correlations between output dimensions.

**Acquisition Function**

Acquisition functions are the core machinery by which subsequent points are selected to query the true underlying function. While many of these utility functions exist, they all aim to strike a balance between exploration of the input space $\mathcal{X}$ and exploiting prominent subspaces. Out of the variety of such acquisition functions available, this work relies upon the well-established Expected Hypervolume Improvement (EHVI) acquisition function[81] due to the multi-objective optimization problem involving a set of target material properties.

Multi-objective optimization involves the simultaneous optimization of multiple conflicting objectives. A common goal is to approximate the Pareto front, which represents the set of non-dominated solutions. In the context of Bayesian optimization, the EHVI acquisition function is a widely used criterion for guiding the selection of which candidate points to evaluate. EHVI balances exploration and exploitation by quantifying the expected improvement in the hypervolume metric, a measure of Pareto front quality.

The hypervolume of a set of points in the objective space is defined as the volume of the region dominated by those points and bounded by a reference point. Let $P$ denote the current Pareto front and $r$ a reference point in the objective space. The hypervolume of $P$ is given by:

$$HV(P) = Volume\left(\bigcup_{p \in P}[p, r]\right) \qquad (7)$$

where $[p, r]$ denotes the hyper-rectangle spanned between $p$ and $r$. A measure of improvement in Hypervolume (HVI) then quantifies the increase in hypervolume achieved by adding a new candidate point $y$ to the Pareto front:

$$HVI(y, P) = max(0, HV(P \cup \{y\}) - HV(P)) \qquad (8)$$

We can then extend this notion to the creation of the EHVI acquisition function, which evaluates the expected value of the HVI under the predictive distribution of the surrogate model. Let $Y$ be the random vector representing the predicted objective values at a candidate input $x$. The EHVI is defined as:

$$EHVI(y, P) = \mathbb{E}[HVI(Y, P)] \qquad (9)$$

where the expectation is taken with respect to the posterior distribution of $Y$ conditioned on the observed data. The computation of EHVI generally requires an analytically intractable integration over the multi-objective posterior distribution, frequently performed via monte-carlo integration.

The q-Expected Hypervolume Improvement (q-EHVI) is an extension of the Expected Hypervolume Improvement (EHVI) that enables evaluation of the EHVI across a batch of q candidate points simultaneously. It measures the expected increase in hypervolume when all q candidates are jointly evaluated, incorporating correlations between their predicted objective values. Formally, q-EHVI is defined as:

$$\alpha_{q-EHVI}(X_{cand}) = \mathbb{E}[HVI(f(X_{cand}))] = \int_{-\infty}^{\infty} HVI(f(X_{cand})) p(f(X_{cand})|D) df \qquad (10)$$

where $X_{cand} = \{x_1, \dots, x_q\}$ is the set of q candidate points, $f(X_{cand})$ are the corresponding objective values, and $p(f(X_{cand})|D)$ is the joint posterior predictive distribution of the model conditioned on the observed data $D$.

Since there is no closed-form solution for $q > 1$ or when the predicted outcomes are correlated, the expectation is approximated using Monte Carlo (MC) integration. This involves drawing $N$ samples $\{f(X_{cand})\}_{t=1}^{N}$ from the joint posterior $p(f(X_{cand})|D)$. Letting $z_{k,X_j,t}^{(m)} = min[u_k, \min_{x' \in X_j} f_t(x')]$, the expectation can be estimated as:

$$\hat{\alpha}_{q-EHVI}(X_{cand}) = \frac{1}{N}\sum_{t=1}^{N} HVI(f_t(X_{cand})) \qquad (11)$$

$$\hat{\alpha}_{q-EHVI}(X_{cand}) = \frac{1}{N}\sum_{t=1}^{N}\sum_{k=1}^{K}\sum_{j=1}^{q}\sum_{X_j \in X_j}(-1)^{j+1} \prod_{m=1}^{M}\left[z_k^{(m)} - l_k^{(m)}\right] \qquad (12)$$

The integration region is divided into $K$ hyper-rectangular cells based on the current Pareto where $z_k^{(m)}$ is the upper bound of the $m$-th objective in the $k$-th cell, and $l_k^{(m)}$ is the lower bound. The overall q-EHVI is obtained by summing the contributions of all active cells and accounting for the combinatorial subsets of the q candidates.

The MC estimation error decreases as $O(1/\sqrt{N})$ with independent samples, regardless of the dimensionality of the search space. To improve efficiency, randomized quasi-Monte Carlo (QMC) methods are often employed, which reduce variance and provide faster convergence in practice.

**FIGURE LEGENDS**

**Figure 1:** Low-dimensional parametrization of a coarse-grained polymer chain.

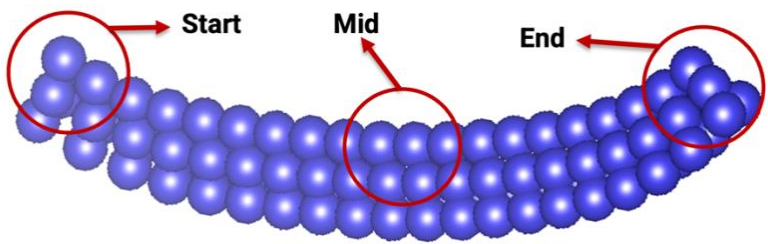

The schematic illustrates the selection of start, middle, and end segments of a 20-polystyrene (20-PS) chain for parameter optimization. This approach reduces the dimensionality of the optimization space from 139 total bonded parameters to 15, focusing on the most influential regions of the polymer.

**Figure 2: The Bayesian optimization workflow for refining coarse-grained topologies.**

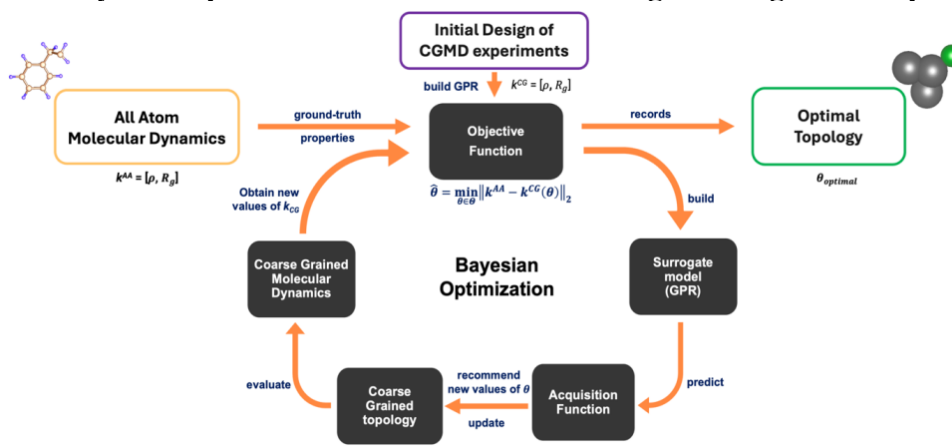

This flowchart illustrates the iterative process where properties from all-atom molecular dynamics (AAMD) serve as the ground truth. A Gaussian Process Regression (GPR) surrogate model and an acquisition function are used to intelligently select new coarse-grained (CG) topology parameters to evaluate, progressively minimizing the difference between CG and AA properties to find an optimal topology.

**Figure 3: Convergence of the Bayesian optimization for macroscopic properties.**

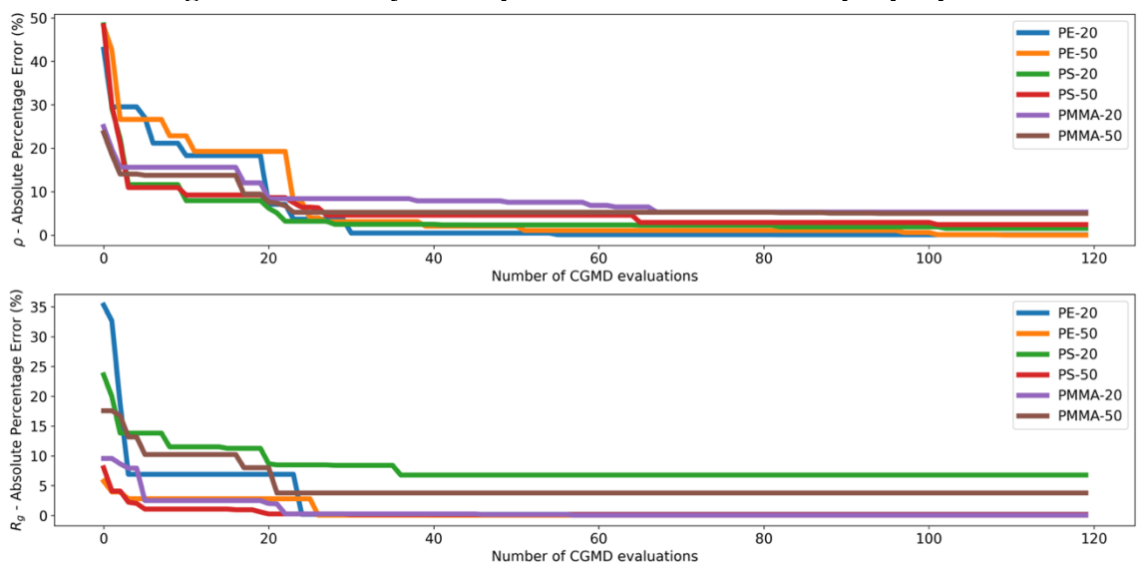

The plots show the absolute percentage error relative to all-atom computations as a function of the number of CGMD evaluations for a) density ($\rho$) and b) radius of gyration ($R_g$). The results for multiple polymer systems demonstrate rapid convergence, typically reaching a plateau within 50 iterations (100 CGMD evaluations).

**Figure 4:** Pareto fronts from multi-objective Bayesian optimization

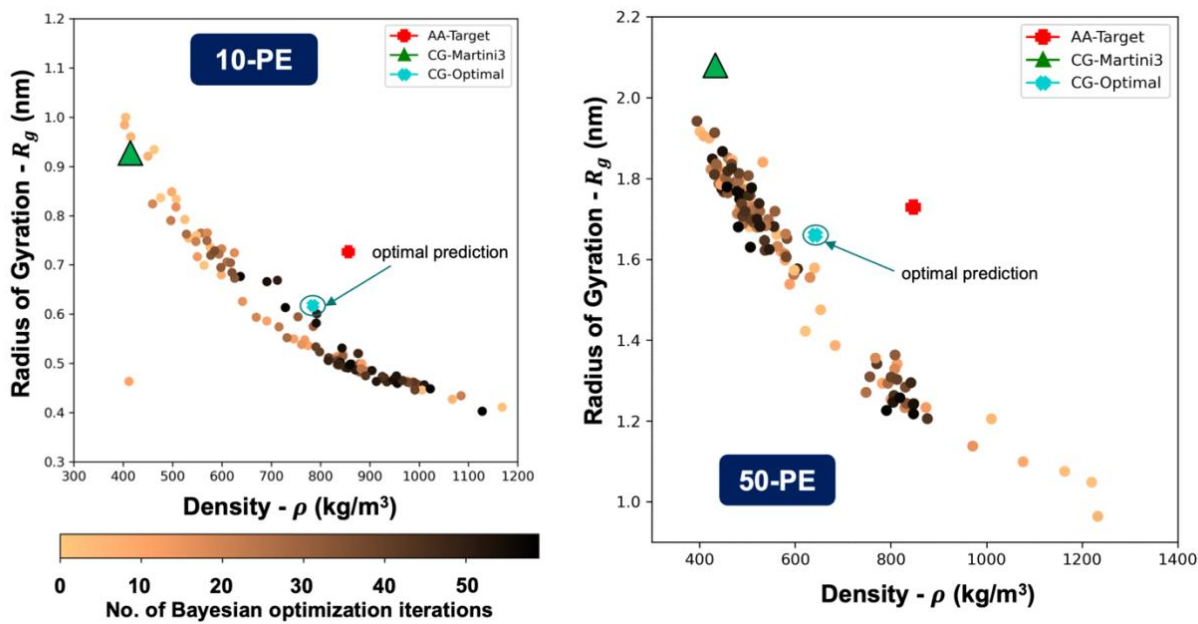

The plots show the evolution of the Pareto front for 10-PE and 50-PE over 60 optimization iterations (including 10 iterations for initial design). Each point represents a CG parameter set, plotted by its resulting density ($\rho$) and radius of gyration ($R_g$). The color of the points indicates the iteration number. The AA target (red circle), initial Martini3 value (green triangle), and an optimal prediction from the final front (cyan circle) are highlighted.

**Figure 5**: Performance comparison of optimized topologies against the default Martini3 model.

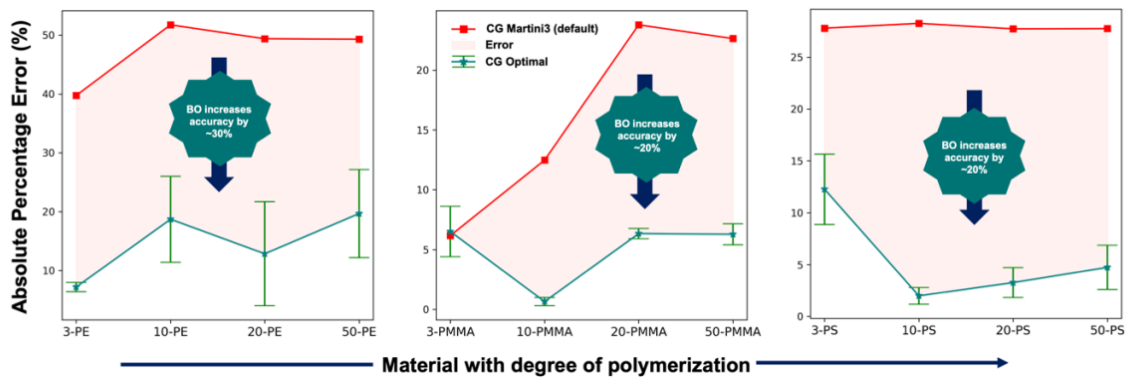

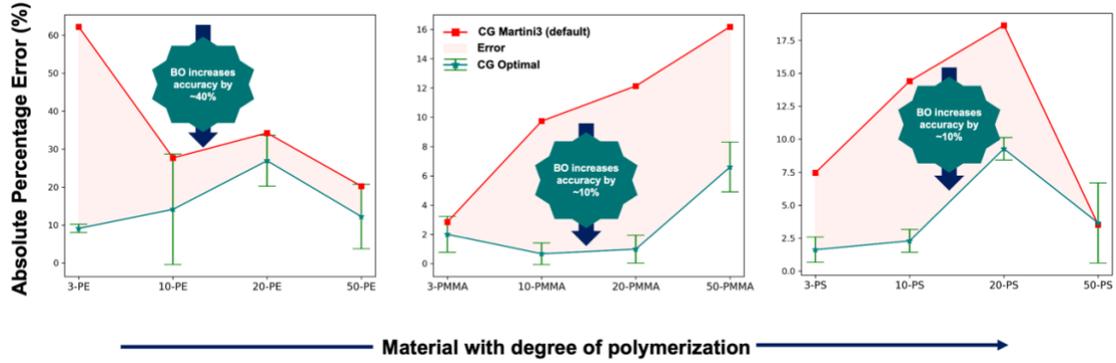

Absolute percentage error for density ($\rho$) and radius of gyration ($R_g$). Each plot compares the error of the default CG-Martini3 model (red line) with the BO-optimized model (green line) across all 12 polymer systems. The shaded area highlights the significant improvement in accuracy achieved by the optimization framework. Error bars represent the standard deviation over five independent runs.

**TABLE LEGENDS**

| Constraints for $\theta$ | PE | PMMA | PS |
|---|---|---|---|
| **Bond Length ($b_0$)** | $0.3 \leq b_0 \leq 0.5$ | Start and End bonds: $0.3 \leq b_0 \leq 0.4$ Middle bonds: $0.26 \leq b_0 \leq 0.36$ | Start and End bonds: $0.22 \leq b_0 \leq 0.32$ Middle bonds: $0.24 \leq b_0 \leq 0.34$ |
| **Bond Constant ($k_b$)** | $1600 \leq k_b \leq 2400$ | Start and End bonds: $8000 \leq k_b \leq 10000$ Middle bonds: $3000 \leq k_b \leq 5000$ | Start and End bonds: $7000 \leq k_b \leq 90000$ Middle bonds: $9000 \leq k_b \leq 11000$ |
| **Angle Magnitude ($\Phi$)** | $167.5 \leq \Phi \leq 172.5$ | Start and End bonds: $67.5 \leq \Phi \leq 72.5$ Middle bonds: $113.5 \leq \Phi \leq 117.5$ | Start and End bonds: $133.5 \leq \Phi \leq 138.5$ Middle bonds: $113.5 \leq \Phi \leq 117.5$ |
| **Angle Constant ($k_\Phi$)** | $10 \leq k_b \leq 15$ | Start and End bonds: $18 \leq k_\Phi \leq 22$ Middle bonds: $30 \leq k_\Phi \leq 40$ | Start and End bonds: $95 \leq k_\Phi \leq 105$ Middle bonds: $30 \leq k_\Phi \leq 40$ |
| **Aromatic Bond Length ($c$)** | ---- | ---- | $0.24 \leq c \leq 0.34$ |
| **Dimensions of Optimization Space** | 12 | 12 | 15 |

**Table 1:** Optimization space (constraints) for the low-dimensional parameters of the coarse-grained polymer systems.

| Polymer | AA Structure (n=20) | Martini3[21,23] Mapping | CG Structure (n=20) |
|---|---|---|---|
| **20-PE** | 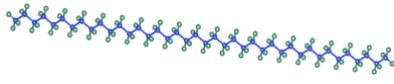 | 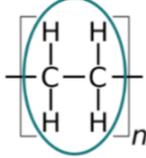 | 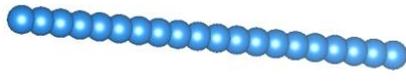 |
| **20-PMMA** | 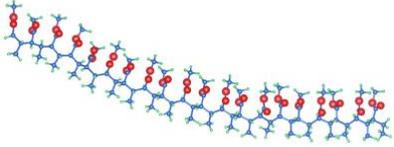 | 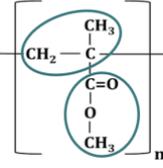 | 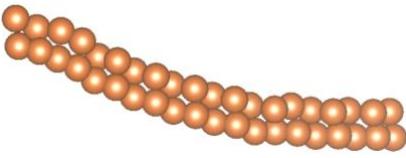 |
| **20-PS** | 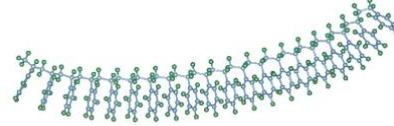 | 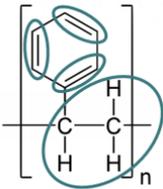 | 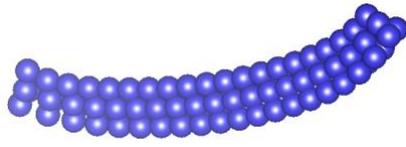 |

**Table 2:** All-atom structure to Coarse-grained structure mapping using Martini3[21,23]